\documentstyle[12pt,epsfig]{article}
\evensidemargin=\oddsidemargin
\def\eps{\epsilon^{\mu \nu \alpha \beta}}

\def\rar{\rightarrow}

\begin{document}
\title{Chiral anomaly and $\eta-\eta^{\prime}$ mixing$^{*}$}
\author{E.P. Venugopal and Barry R. Holstein\\
Department of Physics and Astronomy\\
University of Massachusetts\\
Amherst, MA  01003}
\maketitle
\begin{abstract}
We determine the $\eta-\eta^{\prime}$ mixing angle via a procedure
relatively independent of theoretical assumptions by simultaneously
fitting $\eta,\eta^{\prime}$ reactions involving the 
anomaly---$\eta,\eta^{\prime} \rar \gamma\gamma,~\pi^+\pi^-\gamma$. We extract
reasonably precise renormalized values of the octet and singlet
pseudoscalar decay constants $F_8$, $F_0$, as well as the mixing angle
$\theta$.
\end{abstract}
\vfill
\footnoterule
\leftline{* Research supported in part by the National Science Foundation}
\newpage

\section{Introduction}
From a strictly theoretical perspective, there exists a significant
difference between the octet pseudoscalar
mesons---$\pi$,K,$\eta_8$---and 
their singlet counterpart - $\eta_0$.\cite{33} The former are legitimate
pseudo-Goldstone bosons whose masses vanish in the chiral limit while
the latter is not due to the anomalous breaking of the axial U(3)
symmetry down to SU(3). However, in the real world, this does not seem
to make much difference. Indeed, the physical 
eigenstates---$\eta,\eta^{\prime}$---are 
mixtures of octet and singlet components
\begin{eqnarray}
\eta         &=& \eta_8\cos\theta - \eta_0\sin\theta \nonumber \\
\eta^{\prime}&=& \eta_8\sin\theta + \eta_0\cos\theta
\end{eqnarray}
and the mixing angle $\theta$ is an important quantity in confronting
theoretical calculations with experimental results in these systems.\cite{2}

The mixing angle can be evaluated in various ways but a standard
procedure involves the diagonalization of the $\eta, \eta^{\prime}$ mass
matrix, which at lowest order yields a mixing angle $\theta \approx
-10^{\circ}$. This can easily be seen by writing the mass matrix as
\begin{equation}
m^2 = \left( \begin{array}{cc}
                m^{2}_8  &  m^{2}_{08} \\
                m^{2}_{08}      &  m^{2}_0
                \end{array}
	\right)
\end{equation}
Employing the Gell-Mann-Okubo (GMO) relation to fix $m^{2}_8$ as\cite{34}
\begin{equation}
m^{2}_8 = \frac{1}{3}(4m^{2}_{K} - m^{2}_{\pi})
\label{eq5}\end{equation}
and diagonalizing, one can determine $m^{2}_{08}$, $m^{2}_0$ and
$\theta$, in terms of $m_{\eta},m_{\eta^{\prime}}$, yielding
\begin{equation}
\theta = -9.4^{\circ}, \quad m^{2}_{08} = -0.44m^{2}_{K}, \quad
m_{\eta_0} = 0.95~GeV
\end{equation}

However, the GMO relation, Eq.\ref{eq5}, is valid only at lowest 
order---${\cal O}(p^2$)---in chiral perturbation theory. 
The inclusion of ${\cal O}(p^4$)
corrections to the relation results in significant changes.
Characterizing these via
\begin{equation}
m^{2}_8 = \frac{1}{3}(4m^{2}_{K} - m^{2}_{\pi})(1+\delta)
\end{equation}
a leading logarithm estimate\cite{4} 
\begin{equation}
\delta\approx {-2{m_K^4\over (4\pi F_\pi)^2}\ln{m_K^2\over (4\pi
F_\pi)^2}\over 4m_K^2-m_\pi^2}\label{eq:aa}
\end{equation}
yields $\delta \approx 0.16$, which yields 
\begin{equation}
\theta \cong -20^{\circ}, \quad m^2_{08}=-0.81m_K^2, \quad m_0 = 0.90 GeV 
\end{equation}
suggesting a doubling of the mixing angle. A full one-loop chiral
perturbation theory calculation confirms this finding.\cite{14}
It is also interesting 
that this solution is consistent with the assumptions of
simple $U(3)$ symmetry wherein $\eta_8, \eta_0$ have the same
wavefunction, leading to
\begin{equation}
m^{2}_{08} \cong \frac{2\sqrt2}{3}\left(\frac{\hat{m}-m_s}{\hat{m}+m_s}
\right) \simeq -0.9~m^{2}_{K}
\end{equation}
Note that this result is strongly dependent upon 
chiral symmetry breaking effects on the GMO prediction.

An alternative and independent approach to the problem involves the 
use of the chiral anomaly, which is responsible for the well-known
$\pi,\eta,\eta^{\prime} \rar \gamma\gamma$ decays.\cite{5} 
In the case of 
$\pi^0 \rar \gamma\gamma$, at lowest order---${\cal O}(p^4)$---the
anomalous (Wess-Zumino-Witten) chiral lagrangian predicts\cite{6}
\begin{equation}
{A_{\pi^0 \rar \gamma\gamma}} =  \frac{\alpha N_{c}}{3\pi\bar{F}}
~\epsilon^{\mu \nu \alpha \beta}\epsilon_{1\mu}\epsilon_{2\nu}
k_{1\alpha}k_{2\beta}
\end{equation}
where $\epsilon_{i},k_{i}$ are the polarization, momenta of the
outgoing photons, and $\bar{F}$ is the pion decay constant in the chiral
limit. A leading log calculation of the chiral corrections reveals
that the dominant effect is simply to replace $\bar{F}$ by its physical
value $F_{\pi}=92.4 MeV.\cite{8}$ The resulting amplitude is 
guaranteed by general theorems to remain unchanged in higher chiral
orders.\cite{10}  One then finds that 
the predicted amplitude 
\begin{equation}
F_{\pi\gamma\gamma}(0) = \frac{\alpha N_{c}}{3\pi F_{\pi}} = 0.025~GeV^{-1}
\end{equation}
is in excellent agreement with the corresponding experimental value\cite{12} 
\begin{equation}
F_{\pi\gamma\gamma}(0) = (0.024 \pm 0.001)~GeV^{-1}
\end{equation}
thus providing the confidence that one may be able to analyze the 
$\eta, \eta^{\prime}$ decays with a similar precision. 

In case of the $\eta,\eta^{\prime} \rar \gamma\gamma$ decays, one must
include both, mixing as well as renormalization of the 
octet, singlet couplings, yielding the predicted amplitudes
 \footnote{Note that we implicitly assume here, as do other workers,
that all ${\cal O}(m_{\eta}^{2}, m_{\eta^{\prime}}^{2}/
\Lambda_{\chi}^{2})$ effects,
where $\Lambda_{\chi} \sim 4 \pi F_{\pi}$ is the chiral scale,
are included in the renormalization of the pseudoscalar couplings
$F_8$, $F_0$ and in the mixing angle $\theta$. This does not have to be
the case, but appears to be borne out by the consistency of the results
obtained from treatments of differing manifestations of the anomaly,
as we show below and as others have found.}
\begin{eqnarray}
F_{\eta\gamma\gamma}(0)            &=& \frac{\alpha
N_{c}}{3\sqrt{3}\pi F_{\pi}} 
\left(\frac{F_{\pi}}{F_8}\cos{\theta} - 
2\sqrt{2}\frac{F_{\pi}}{F_0}\sin{\theta}\right) \nonumber \\
F_{\eta^{\prime}\gamma\gamma}(0)   &=& \frac{\alpha
N_{c}}{3\sqrt{3}\pi F_{\pi}}
\left(\frac{F_{\pi}}{F_8}\sin{\theta} + 
2\sqrt{2}\frac{F_{\pi}}{F_0}\cos{\theta}\right)
\end{eqnarray}
From the PDG one extracts experimental values\cite{12}
\begin{eqnarray}
F_{\eta\gamma\gamma}(0)          &=& 0.0249 \pm 0.0010~GeV^{-1}\nonumber \\
F_{\eta^{\prime}\gamma\gamma}(0) &=& 0.0328 \pm 0.0024~GeV^{-1}
\end{eqnarray}

In order to solve the system, however, we require an additional
assumption since there are three unknowns---$F_8, F_0, \theta$---and
only two pieces of data. The usual approach in this case is to use
the leading log prediction 
of one-loop chiral perturbation theory to predict\cite{14}
\begin{equation}
{F_8\over F_\pi}  = \left[1 - \frac{m_K^2}{(4\pi F_{\pi})^2}~
\ln{\frac{m_K^2}{(4\pi F_\pi)^2}}+\frac{m_\pi^2}{(4\pi
F_\pi)^2}\right] 
\approx 1.30 
\label{eq15}\end{equation}
and then solve for $F_0, \theta$, yielding
\begin{equation}
\frac{F_0}{F_\pi} = 1.04, \quad \theta = -20^{\circ}
\end{equation}
It is interesting to note that these results are quite consistent with those
obtained from the one-loop analysis of the mass matrix---{\it i.e.} $\theta
\approx -20^{\circ}$ and $F_0/F_\pi$ consistent with the
value---unity---which one would
have if the singlet state and the pion were to have the same wavefunction.

While this agreement is satisfying, the extraction of these mixing
parameters requires certain theoretical inputs, either Eq. \ref{eq15}
or Eq. \ref{eq:aa}, and it is interesting
to inquire whether one can predict the mixing angle purely
phenomenologically. As we shall show, the answer is yes, provided one
utilizes the additional information available from the anomalous
decays  $\eta,\eta^{\prime} \rar \pi^+\pi^-\gamma$.\cite{100}

In the next section then we show how the decays 
$\eta,\eta^{\prime} \rar \pi^+\pi^-\gamma$ can be analyzed in order to
isolate the chiral anomaly. This involves a careful study of final
state interactions and unitarity constraints in order to realistically
extrapolate to zero four-momenta, as required by the anomaly. In the
concluding section, we apply these results to evaluate $\theta, F_8,
F_0$ in an essentially model independent fashion.

\section{Analysis of $\eta,\eta^{\prime} \rar \pi^+\pi^-\gamma$ Decays}

Our goal is to use the experimental data on 
$\eta,\eta^{\prime} \rar \pi^+\pi^-\gamma$ in order to isolate
the value of the anomaly in these decays.  The resulting numbers
can then be fit to the
appropriate theoretical expressions, thus allowing extraction of
the renormalized mixing angle and coupling constants. To this end, we
define
\begin{equation}
A_{\eta,\eta^{\prime} \rar \pi\pi\gamma} = 
B_{\eta,\eta^{\prime}}(s,s_{\pi\pi})~\eps
\epsilon_{\mu}k_{\nu}p_{+\alpha}p_{-\beta}\\
\end{equation}
where $p_{\pm}$, k are the outgoing 4-momenta, $\epsilon_{\mu}$ is the photon
polarization vector, $s=m_{\eta,\eta^{\prime}}^2$ and
$s_{\pi\pi}=(p_++p_-)^2$. The chiral anomaly ({\it cf.} Figure 1(a)) requires
\begin{eqnarray}
B_{\eta}(0,0)= \frac{eN_c}{12 \sqrt{3} \pi^{2} F_\pi^{3}}
\left(\frac{F_\pi}{F_8}\cos{\theta} -
\sqrt{2}\frac{F_\pi}{F_0}\sin{\theta}
\right)\nonumber \\
B_{\eta^{\prime}}(0,0)= \frac{eN_c}{12 \sqrt{3} \pi^{2} F_\pi^{3}}
\left(\frac{F_\pi}{F_8}\sin{\theta} +
\sqrt{2}\frac{F_\pi}{F_0}\cos{\theta}
\right)\label{eq:ww}
\end{eqnarray}

\begin{figure}[htb]
\centering
\leavevmode
\centerline{\epsfbox{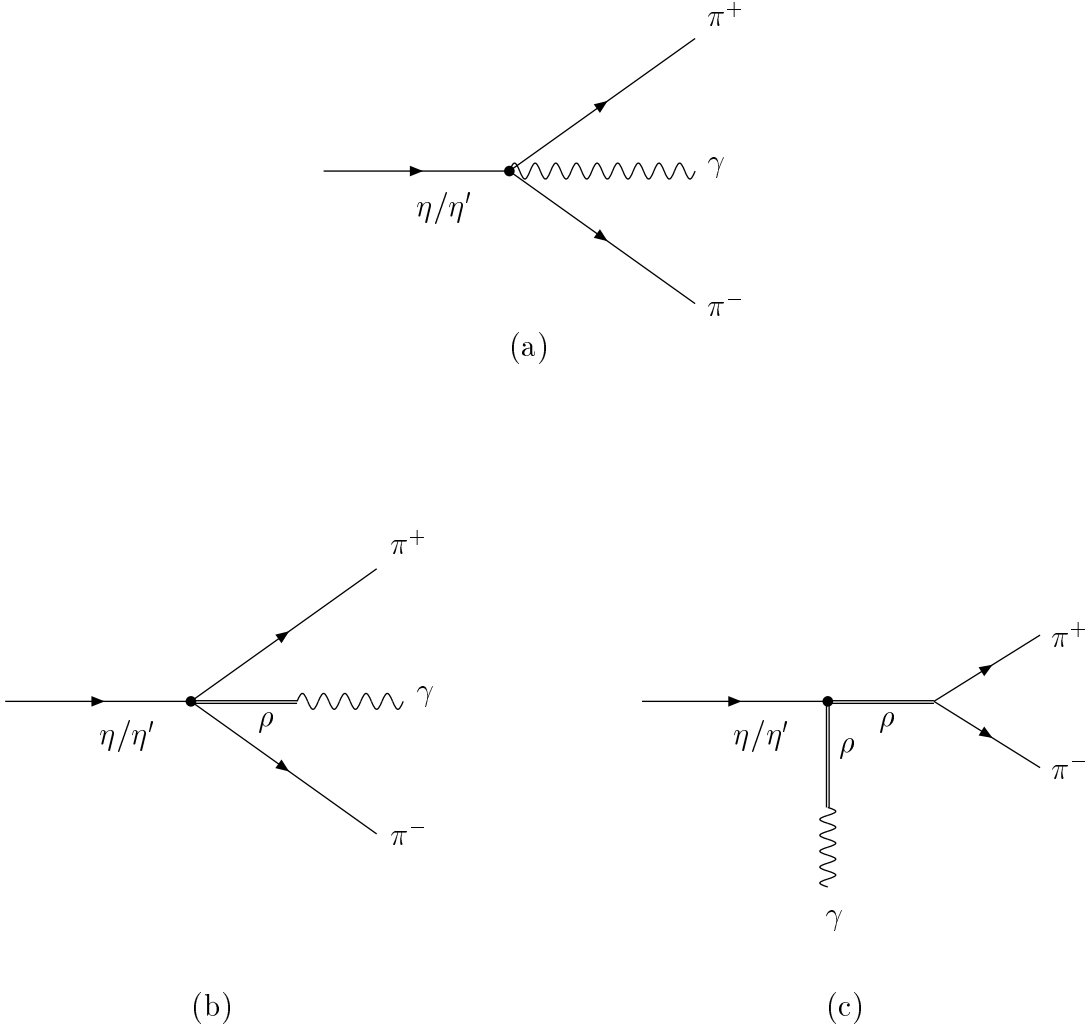}}
\caption{Shown are contact (a) and VMD (b,c) contributions to
$\eta,\eta'\rightarrow\pi^+\pi^-\gamma$ decay.}
\label{fig1}\end{figure}

Note, however, that the chiral anomaly, strictly
speaking, only constrains the form factors
$B_{\eta,\eta^{\prime}}(s,s_{\pi\pi})$ at zero four 
momentum---$B_{\eta,\eta^{\prime}}(0,0)$---while 
the experimental input occurs at
$s=m_{\eta,\eta^{\prime}}^{2}$, $s_{\pi\pi} \geq 4m_{\pi}^{2}$.  One
indication of this fact is that, using the simple energy-independent 
form given in 
Eq. \ref{eq:ww} to calculate the decay rate, one obtains for the
$\eta$-channel the value $\Gamma_{\eta-\pi\pi\gamma}=35.7$
eV compared to the experimental rate of 
$\Gamma_{\eta-\pi\pi\gamma}^{\rm exp}=64\pm 6$ eV.  For the $\eta'$
channel things are even worse---$\Gamma_{\eta'-\pi\pi\gamma}=61\pm 5$
KeV while the theoretical value arising from use of the simple anomaly
amplitude is a factor of twenty less!  We conclude that in order to
extract values for the anomaly in these transitions, it is absolutely
essential to correctly model the energy dependence of the amplitude in the
physical region. 

This problem is not unique, of course, to the $\eta,\eta^{\prime} \rar \pi\pi\gamma$
system and has been addressed previously in the extraction of the anomaly 
in $\gamma \rar 3\pi$ from the Primakoff-effect data of Antipov et 
al.\cite{16} In this case, one also has a clear prediction of the 
chiral anomaly.\cite{18} Writing
\begin{equation}
A_{\gamma \rar 3\pi} = F_{3\pi}(s,t,u)~\eps
\epsilon_{\mu}p_{+\nu}p_{-\alpha}p_{0\beta}
\end{equation}
where $s=(p_++p_-)^2, t=(p_++p_0)^2, u=(p_-+p_0)^2$, the chiral stricture demands
\begin{equation}
F_{3\pi}(0,0,0) = \frac{eN_c}{12\pi F^{3}_{\pi}} = 9.7~GeV^{-3}
\end{equation}
The effects of p-wave pi-pi interactions at low and moderate energies
are known to be reasonably described by vector dominance.\cite{20}
In particular, the models described in ref.\cite{22}, which incorporate vector
dominance and chiral symmetry, when applied to the $\gamma \rar 3\pi$ 
reaction, provide a form
\begin{equation}
F_{3\pi}(s,t,u) = -\frac{1}{2}F_{3\pi}(0,0,0)\left[1 -
\frac{m^{2}_{\rho}}{m^{2}_{\rho}-s} -
\frac{m^{2}_{\rho}}{m^{2}_{\rho}-t} -
\frac{m^{2}_{\rho}}{m^{2}_{\rho}-u}\right]
\end{equation}
which matches on to the anomaly at zero four-momentum but also offers a
plausible extension into the physical region.  
This form must be modified, of course, in order to confront real data
since unitarity demands the presence of branch cuts and consequent 
imaginary components in the
form factor.  This is clear from a one-loop chiral perturbation theory
calculation, which yields\cite{24}
\begin{equation}
F_{3\pi}(s,t,u) = F_{3\pi}(0,0,0) \left[ 1 + {3m_{\pi}^{2} \over
 2m_{\rho}^{2}} + {m_{\pi}^{2} \over 24\pi^2 F_{\pi}^{2}}
 \left({3\over 4} \ln{m_{\rho}^{2} \over m_{\pi}^{2}} + F(s) + F(t) +
F(u)\right) \right]
\end{equation}
where
\begin{equation}
F(s)=\left\{
\begin{array}{ll}
(1-{s\over 4m_\pi^2})\sqrt{s-4m_\pi^2\over s}\ln{1+\sqrt{s-4m_\pi^2\over
s}\over -1+\sqrt{s-4m_\pi^2\over s}}-2 & s>4m_\pi^2 \\
2(1-{s\over 4m_\pi^2})\sqrt{4m_\pi^2-s\over s}\tan ^{-1}\sqrt{s\over
4m_\pi^2-s}-2& s \leq 4m_\pi^2
\end{array}\right.
\end{equation}
Here we note that the imaginary component of the function $F(s)$ is
given in terms of the energy-dependent width of the rho meson via
\begin{equation}
{m_\pi^2\over 24\pi^2F_\pi^2}\mbox{Im}F(s)={1\over m_\rho} \Gamma_\rho(s)
\end{equation}
where
\begin{equation}
\Gamma_{\rho}(s) = \frac{g_{\rho\pi\pi}^{2}s}{48\pi m_{\rho}}
(1 - \frac{4m_{\pi}^2}{s})^{3/2}
\end{equation}

This one-loop form is no doubt appropriate in the near threshold region.
However, once $s_{\pi\pi  } \geq 10 m^{2}_{\pi}$ or so, one does not
anticipate that a simple one-loop description will be adequate. In 
order to address these difficulties, the use of an N/D form has
been suggested.\cite{77}  In this approach, one utilizes the
Omnes function, which encodes information concerning the pi-pi 
interaction\cite{26}
\begin{equation}
D_1(s) = exp\left(-\frac{s}{\pi}~\int_{4m_{\pi}^{2}}^{\infty}~
\frac{ds^{\prime}~\delta_1(s^{\prime})}{s^{\prime}(s^{\prime}
-s-i\epsilon)}\right)\label{eq:jj}
\end{equation}
where $\delta_1(s)$ is the ($l=1$) p-wave pi-pi phase shift at center
of mass energy $\sqrt{s}$.  There are two ways to proceed at this
point.

\begin{itemize}  

\item [i)] One can use the experimental phase shifts, with some
assumptions made about their asyptotic form.  In our case we took the
values quoted by Froggatt and Peterson,\cite{78} which are given up to
$\sqrt{s}$ = 1 GeV and assumed a constant value after that.  We label
the Omnes function obtained in this way as $D_1^{\rm exp}(s)$.

\item [ii)] One can employ a simple analytic form\cite{28}
\begin{equation}
D_1(s) = 1-\frac{s}{m_{\rho}^{2}}-\frac{s}{96\pi^2 F_{\pi}^2}
ln\frac{m_{\rho}^{2}}{m_{\pi}^{2}}-\frac{m_{\pi}^{2}}{24\pi^2F_{\pi}^2}
F(s)\label{eq:gg}
\end{equation}
which has been shown to provide an approximate description of the 
empirical $\pi\pi$ p-wave phase shifts in the low energy
region.\cite{30}  We label the Omnes function obtained via this
procedure as $D_1^{\rm anal}(s)$.

\end{itemize}

Using either of these forms, and postulating
an N/D form of the $\gamma \rar 3\pi$ amplitude as
\begin{eqnarray}
F_{3\pi}(s,t,u)&=&-{1\over 2}A_{3\pi}(0)[1-({m_\rho^2\over
m_\rho^2-s}+{m_\rho^2\over m_\rho^2-t}+{m_\rho^2\over m_\rho^2-u})]\nonumber\\
&\times&\left({1-{s\over m_\rho^2}\over D_1(s)}\right)
\left({1-{t\over m_\rho^2}\over D_1(t)}\right)\left({1-{u\over m_\rho^2}\over
D_1(u)}\right)
\label{eq35}\end{eqnarray}
one can see that Eq. \ref{eq35} matches onto the one-loop chiral result in the
limit of low energies {\it and} onto the vector dominance form when
unitarity inspired logarithms are dropped.

Whether such an N/D form accurately describes the data for $\gamma \rar 3\pi$
awaits the arrival of sufficiently precise information from CEBAF and
CERN. However, it certainly appears to satisfy the various criteria
which nature demands, and suggests the treatment of the related
$\eta,\eta^{\prime} \rar \pi^+\pi^-\gamma$ decay amplitudes in a
parallel fashion.

In the case of the $\eta,\eta'\rightarrow\pi^+\pi^-\gamma$ decays we
can proceed similarly.  In this case, the one-loop chiral perturbation
theory calculation gives
\begin{equation}
B_\eta^{\rm 1-loop}(s,s_{\pi\pi})=B_\eta(0,0)[1+{1\over
32\pi^2F_\pi^2}
((-4m_\pi^2+{1\over3}s_{\pi\pi})\ln{m_\pi^2\over m_\rho^2}+{4\over
3}F(s_{\pi\pi})-{20\over 3}m_\pi^2+{3\over 2m_\rho^2}s_{\pi\pi}]
\end{equation}
while vector dominance ({\it cf.} Figure 1(b,c))yields
\begin{equation}
B_{\eta,\eta'}(s,s_{\pi\pi})=B_{\eta,\eta'}(0,0)[1+{3\over
2}{s_{\pi\pi}\over m_\rho^2-s_{\pi\pi}}]
\end{equation}
Certainly, in order to treat the decay of the $\eta'$, one must go further and
include unitarity effects via final state interactions. One very
simple approach is to include the (energy-dependent) width of the 
rho-meson in the propagator via
\begin{equation}
{s_{\pi\pi}\over m_\rho^2-s_{\pi\pi}}\rightarrow
{s_{\pi\pi}\over m_\rho^2-s_{\pi\pi}-im_\rho\Gamma_\rho(s_{\pi\pi})}
\end{equation}
This use of vector width-modified vector dominance 
already makes an important difference from the simple anomaly---tree 
level---results (especially in the case of
the $\eta'$), changing the predicted decay widths from the values 
35 eV and 3 KeV quoted above to the much more realistic numbers
\begin{equation}
\Gamma_{\eta-\pi\pi\gamma}=62.3 eV, \qquad 
\Gamma_{\eta'-\pi\pi\gamma}=67.5 KeV 
\end{equation}
if the parameters
\begin{equation}
F_8/F_\pi=1.3,\qquad F_0/F_\pi=1.04,\qquad \theta=-20^\circ\label{eq:hh}
\end{equation} 
are employed.  However, this approach
does not reproduce the one-loop chiral form in the low energy limit.

In order to determine a form for the final state interactions which
matches onto both the one-loop chiral correction {\it and} to the vector
dominance result in the appropriate limits, we postulate an N/D
structure, as in the related $\gamma\rightarrow 3\pi$ case---
\begin{equation}
B_{\eta-\pi\pi\gamma}(s,s_{\pi\pi})=B_{\eta-\pi\pi\gamma}(0,0)\left[
1-c+c{1+as_{\pi\pi}\over D_1(s_{\pi\pi})}\right]\label{eq:ii}
\end{equation}
where for the Omnes function we use one of the two forms itemized
above and $a,c$ are free parameters to be determined.  In
order to reproduce the coefficient of the $F(s_{\pi\pi})$ function, which
contains the rho width, we require $c=1$.  On the other hand, matching
onto the VMD result at ${\cal O}(p^6)$ can be achieved by the choice 
$a=1/2m_\rho^2$.  Thus in the case of the $\eta$ the form is
completely determined.  Since the $\eta'$ spectrum is closely related
and is dominated by the presence of the rho we shall assume an
identical form for the $\eta'$ case.  Using these forms we can
then calculate the decay widths assuming the theoretical values for
the anomaly.  Using the parameters given in Eq. \ref{eq:hh}
one finds, for example, 
\begin{eqnarray}
 i)D_1^{\rm exp}(s)  && \Gamma_{\eta-\pi\pi\gamma}=65.7 eV, \quad
\Gamma_{\eta'-\pi\pi\gamma}=66.2 KeV\nonumber\\
ii)D_1^{\rm anal}(s) && \Gamma_{\eta-\pi\pi\gamma}=69.7 eV, \quad
\Gamma_{\eta'-\pi\pi\gamma}=77.8 KeV
\end{eqnarray}
There is a tendency then for the numbers obtained via the analytic
form of the Omnes function to be somewhat too high.  

\begin{figure}[htb]
\leftline{
\begin{minipage}[t]{.47\linewidth}
\mbox{\leftline{\epsfig{file=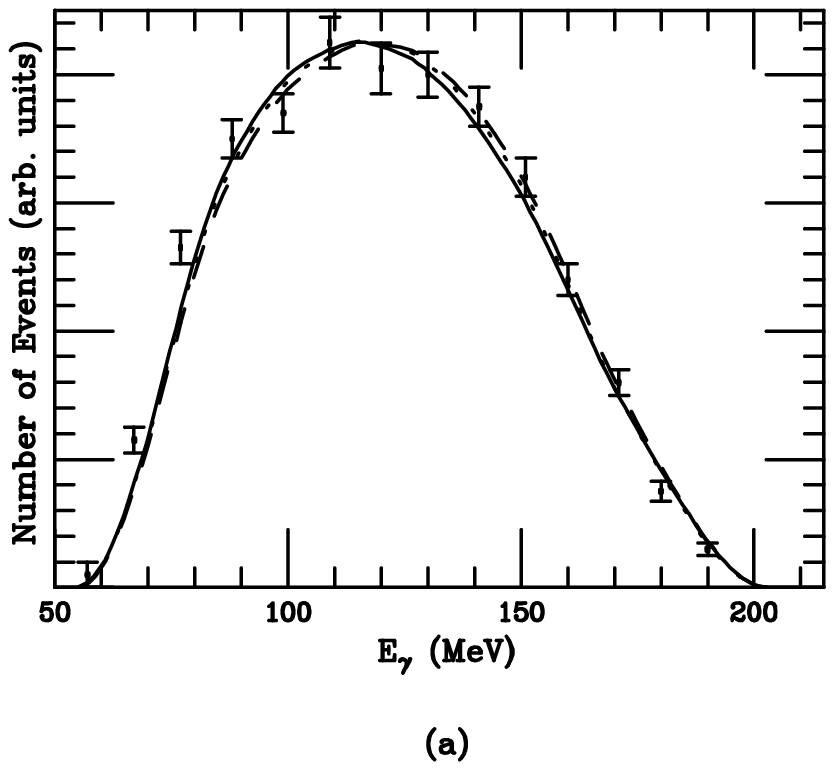,width=7.0cm}}
\rightline{\epsfig{file=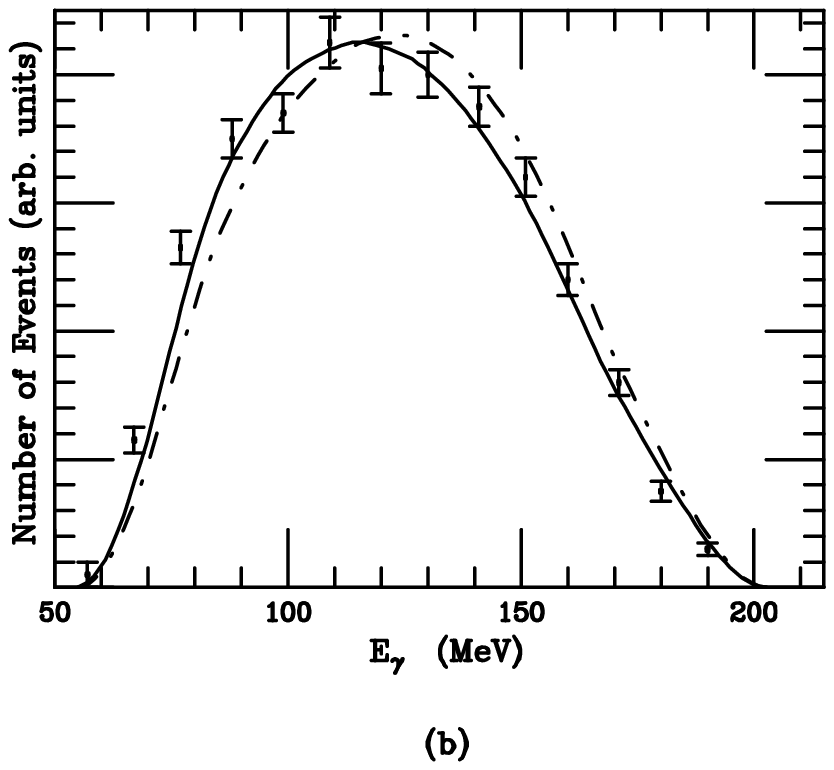,width=7.0cm}}}
\end{minipage}}
\caption{Shown is the photon spectrum in
$\eta\rightarrow\pi^+\pi^-\gamma$ from Gormley et al.\protect\cite{93} as well as
various theoretical fits.  In the first figure, the dashed line represents the
(width-modified) VMD model.  The (hardly visible) dotted line and the
solid line represent the
final state interaction ansatz Eq.\ref{eq:ii} with use of the 
analytic and experimental version of the Omnes function respectively.
The second figure shows the experimental Omnes
function result (solid line) compared with the one-loop result (dotdash line).}
\label{fig2}\end{figure}

We can also
compare the predicted spectra with the corresponding experimentally
determined values.  As shown in Figure 2, we observe that the
experimental spectra are well fit in the $\eta$ case 
in terms of both the N/D or the VMD
forms, but that the one-loop chiral expression does not provide an
adequate representation of the data.\cite{80}  In the case of the corresponding
$\eta'$ decay the results are shown in Figure 3, wherein we observe
that either the unitarized VMD or the use of $N/D_1^{\rm exp}$ provides
a reasonable fit to the data (we get $\chi^2$/dof=32/17 and 20/17,
respectively), while the use of the analytic form for the Omnes
function yields a predicted spectrum ($\chi^2$/dof=104/17) 
which is slightly too low on the high energy end.  However, for both 
$\eta$ and $\eta'$ we see that our simple ansatz---Eq.\ref{eq:ii}---provides 
a very satisfactory representation of the decay spectrum.

\begin{figure}[htb]
\centering
\leavevmode
\centerline{\epsfbox{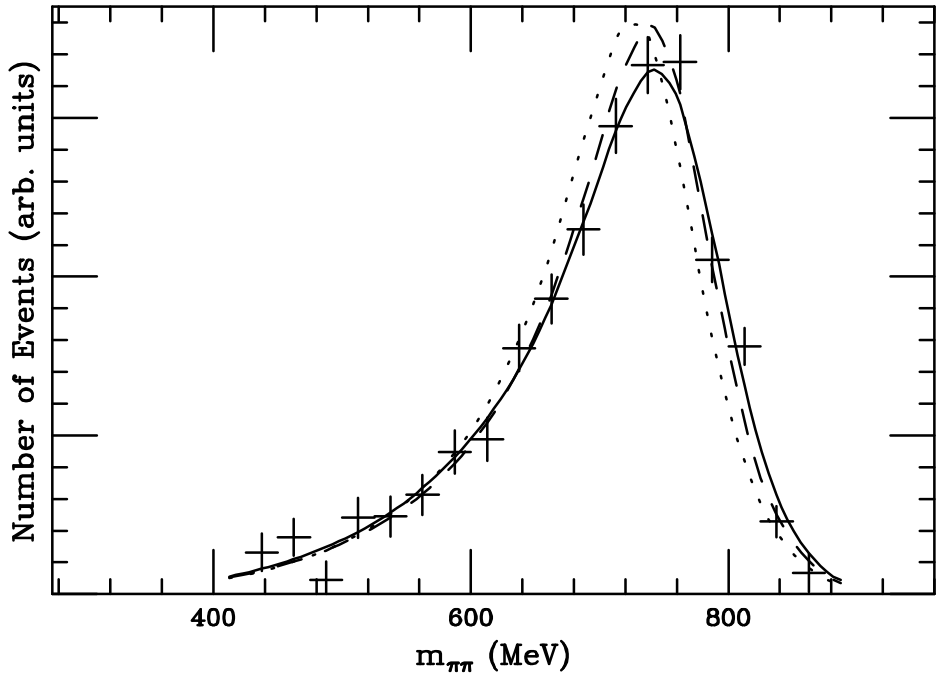}}
\caption{Shown is the photon spectrum in 
$\eta'\rightarrow\pi^+\pi^-\gamma$ from Abele et al.\protect\cite{83} as well
as various theoretical fits.  As in Figure 2, the dashed line represents the
(width-modified) VMD model.  The dotted and solid lines represent 
the final state interaction ansatz Eq.\protect\ref{eq:ii}  with use 
of the analytic and experimental version of the Omnes function, respectively.
Here the curves have been normalized to the same number of events.}
\label{fig3}\end{figure}

\section{Evaluation of $\eta-\eta^{\prime}$ mixing parameters}

Our conclusion in the last section was that if the mixing angle and
pseudoscalar coupling constants were given values consistent with
present theoretical and experimental leanings, then the predicted
widths and spectra of
both $\eta,\eta\rightarrow\pi^+\pi^-\gamma$ are basically consistent with
experimental values.  Our goal in this section is to go the other way,
however.  That is, using the assumed N/D forms for the decay amplitude,
and treating the pseudoscalar decay constants $F_8,F_0$ as well as the
$\eta-\eta'$ mixing angle $\theta$ as free parameters, we wish to
inquire as to
how well they can be constrained purely from the experimental data
on $\eta,\eta'\rightarrow\gamma\gamma$ and
$\eta,\eta'\rightarrow\pi^+\pi^-\gamma$ decays, with reasonable
assumptions made about the final state interaction effects in these
two channels. 

On theoretical grounds, one is somewhat more confident about the
extraction of the threshold amplitude in the case of the lower energy
$\eta\rightarrow\pi^+\pi^-\gamma$ system.  Indeed, in this case the
physical region extends only slightly into the tail of the rho unlike 
the related $\eta'$ decay wherein the spectrum extends completely over the
resonance so that there exists considerable sensitivity to details 
of the shape.  Thus a first approach might be to utilize only the
two-photon decays together with the
$\eta\rightarrow\pi^+\pi^-\gamma$ width in order to determine the
three desired parameters.  In this fashion one finds the results shown
in Table 1.
\begin{table}
\begin{center}
\begin{tabular}{|c|c|c|c|} \hline
 &$F_8/F_\pi$    & $F_0/F_\pi$    & $\theta$\\ \hline
VMD & $1.28\pm 0.24$ & $1.07\pm 0.48$ & $-20.3^\circ\pm 9.0^\circ$\\
N/D$_1^{\rm anal}$ & $1.49\pm 0.29$ & $1.02\pm 0.42$ & 
$-22.6^\circ\pm 9.6^\circ$\\
N/D$_1^{\rm exp}$ & $1.37\pm 0.26$& $1.02\pm 0.45$& $-21.2^\circ\pm
9.3^\circ$\\ \hline
\end{tabular}
\caption{Values of the renormalized pseudoscalar coupling
constants and the $\eta-\eta'$ mixing angle using the
$\eta,\eta'-\gamma\gamma$ and $\eta-\pi\pi\gamma$ amplitudes in a
three parameter fit.}
\end{center}
\end{table}
We observe that the results are in agreement, both with each other and
with the chiral symmetry expectations---$F_8/F_\pi\sim 1.3$,
$F_0/F_\pi\sim 1$, and $\theta\sim -20^\circ$.  However, the
uncertainties obtained in this way are uncomfortably high.  

In order to ameliorate this problem, we have also done a maximum likelihood fit
including the $\eta'-\pi\pi\gamma$ decay rate, yielding the results
shown in Table 2. 
\begin{table}
\begin{center}
\begin{tabular}{|c|c|c|c|} \hline
 &$F_8/F_\pi$& $F_0/F_\pi$& $\theta$\\ \hline
VMD & $1.28\pm 0.20$& $1.07\pm 0.04$ & $-20.8^\circ\pm 3.2^\circ$\\
N/D$_1^{\rm anal}$& $1.48\pm 0.24$& $1.09\pm 0.03$& $-24.0^\circ\pm
3.0^\circ$\\
N/D$_1^{\rm exp}$&$1.38\pm 0.22$&$1.06\pm 0.03$&$-22.0^\circ\pm
3.3^\circ$\\ \hline
\end{tabular}
\caption{Values of the renormalized pseudoscalar coupling
constants and of the $\eta-\eta'$ mixing angle obtained from a maximum
likelihood analysis using the $\eta,\eta'-\gamma\gamma$ and
$\eta,\eta'-\pi\pi\gamma$ amplitudes.}
\end{center}
\end{table}
We observe that the central values stay fixed but that the error bars
are somewhat reduced.  The conclusions are the same,
however---substantial renormalization for $F_8\sim 1.3 F_\pi$, almost
none for $F_0\sim F_\pi$, and a mixing angle $\theta\sim -20^\circ$.
These numbers appear nearly invariant, regardless of the approach.

\section{Conclusion}

Before summarizing the results of our above analysis, it should
certainly be emphasized that we are not the first to undertake the
program of isolating the anomaly from the $\eta,\eta'-\pi\pi\gamma$
data.  Indeed, there has been considerable work in this regard, both on
the theoretical side \cite{100,82} as well as experimentally, 
including the most recently published $\eta'\rightarrow\pi^+\pi^-\gamma$
data.\cite{83} The recent analysis of ref.\cite{82} leads to results 
quite different from ours in both the mixing angle as well as the 
renormalization of the pseudoscalar couplings. On the other hand, 
ref.\cite{83} (at least for the model labeled $M_1$) finds a somewhat 
smaller mixing angle ($\theta\sim -16^\circ$) and pseudoscalar 
renormalization ($F_8/F_0\sim 1.1$).

However, there is an important difference between these analyses and
our own. In refs.\cite{82} and \cite{83}, the decay amplitude is
written in terms of a piece due to the anomaly (parametrized by
$E_X, X=\eta,\eta'$) and a component due to the rho pole 
(parametrized by $F_X, X=\eta,\eta'$). The ratio $E_X/F_X$ is then
fitted, through a minimization procedure, to produce the
experimental spectrum and partial widths. In our analysis, 
the parameters of the two pieces are fixed ${\it a~priori}$ to 
reproduce the results of one-loop
chiral perturbation theory \cite{24} (by fixing $c=1$) and 
VMD \cite{22} (by fixing $a=1/2m_{\rho}^2$). 
(Indeed a recent analysis of other I=1 $\pi^+\pi^-$
processes found that only within a model such as ref.\cite{22}, which
links chiral symmetry and VMD, could the data be fit
consistently \cite{92}). However, following the picture of 
ref.\cite{22}, we do not include a non-resonant coupling in 
$\gamma\pi^+\pi^-$ (as considered in models $M_1$ and $M_3$ of \cite{82}). 
Our successful fits of the experimental data speak for themselves.

In previous treatments of the $\eta,\eta'$ system via the anomaly, 
which have omitted $\eta,\eta'-\pi\pi\gamma$ constraints, 
the mixing angle $\theta$ has generally been determined only at the 
cost of theoretical assumptions about the renormalization of the octet
pseudoscalar coupling constant $F_8$ with respect to $F_\pi$.  We have in
this paper asked whether it is possible to obtain the mixing angle in
a fashion relatively independent of such theoretical assumptions by
simultaneously fitting $\eta,\eta'\rightarrow\gamma\gamma$ as well as 
$\eta,\eta'\rightarrow\pi^+\pi^-\gamma$ decays.  As shown above, the
answer is affirmative.  However, one must incorporate {\it some} sort 
of model for the final state interactions of the pi-pi system in order to
extrapolate down to zero four-momentum where the anomaly obtains.  
We have argued that the $N/D_1^{\rm exp}$ form given in
Eq. \ref{eq:ii} is reasonable both on theoretical grounds---matching
both the requirements of VMD and of low energy chiral symmetry---and via
successful fitting of the experimental spectra.  Using this form and the PDG
values for $\eta,\eta'\rightarrow\gamma\gamma$ and 
$\eta,\eta'\rightarrow\pi\pi\gamma$ amplitudes we have obtained values
\begin{equation}
F_8/F_\pi= 1.38\pm 0.22\qquad F_0/F_\pi= 1.06\pm 0.03\qquad 
\theta=-22.0^\circ\pm 3.3^\circ\label{eq:qq}
\end{equation}
which are quite consistent with those obtained in previous analyses
which required assumptions about chiral symmetry breaking.  One can
then assess these results in two different ways.  Although it is our
contention that the assumptions made above concerning pion-pion
interactions are relatively model-independent and that the numbers
given thereby in Eq. \ref{eq:qq} are quite solid, one could
also take a contrary view that the forms utilized for final state
interactions {\it do} require critical dynamical assumptions.  In this case,
however, we would argue that via three quite different routes
\begin{itemize}
\item[i)] mass matrix analysis including GMO breaking
\item[ii)] $\eta,\eta'\rightarrow\gamma\gamma$ analysis with
assumptions made about $F_8/F_\pi$
\item[iii)] simultaneous analysis of
$\eta,\eta'\rightarrow\gamma\gamma$ and
$\eta,\eta'\rightarrow\pi^+\pi^-\gamma$ with (minimal) assumptions concerning
final state pi-pi interactions
\end{itemize}
one finds virtually the same value of the mixing 
angle---$\theta\simeq -20^\circ$---and for pseudoscalar
couplings---$F_8/F_\pi\sim 1.3,F_0/F_\pi\sim 1.0$.  In any case, 
we would assert that these values are now strongly (and independently)
confirmed from within the chiral anomaly sector.

\end{document}